\documentclass[a4paper]{article}

\usepackage{INTERSPEECH2022}
\usepackage{tabularray}
\usepackage{cite}
\usepackage{amsmath,amssymb,amsfonts}
\usepackage{algorithmic}
\usepackage{graphicx}
\usepackage{textcomp}
\usepackage{xcolor}

\usepackage{array}
\usepackage{booktabs}
\usepackage{multirow}
\usepackage{epsfig}
\usepackage{bm}
\usepackage{algorithm}
\usepackage{diagbox}
\usepackage{amssymb}
\usepackage{makecell}
\usepackage{adjustbox}
\usepackage{enumitem}
\usepackage{marvosym}
\usepackage{braket}
\usepackage{color}
\usepackage{hyperref}

\title{The USTC-NERCSLIP Systems for the CHiME-8 MMCSG Challenge}
\name{Ya Jiang\textsuperscript{1,\dag}, Hongbo Lan\textsuperscript{1,\dag}, Jun Du$^{1,*}$\thanks{\textsuperscript{\dag}Equal contribution \quad * Corresponding author}, Qing Wang$^1$, Shutong Niu$^1$}
\address{
  $^1$University of Science and Technology of China, China}
\email{\{yajiang,lhb1900,niust\}@mail.ustc.edu.cn, \{jundu,qingwang2\}@ustc.edu.cn}

\begin{document}

\maketitle
\begin{abstract}
In the two-person conversation scenario with one wearing smart glasses, transcribing and displaying the speaker's content in real-time is an intriguing application, providing a priori information for subsequent tasks such as translation and comprehension. Meanwhile, multi-modal data captured from the smart glasses is scarce. Therefore, we propose utilizing simulation data with multiple overlap rates and a one-to-one matching training strategy to narrow down the deviation for the model training between real and simulated data. In addition, combining IMU unit data in the model can assist the audio to achieve better real-time speech recognition performance.
\end{abstract}

\section{Introduction}

In recent years, with the development of deep learning, automatic speech recognition (ASR) has made rapid progress~\cite{li2022recent}, especially end-to-end (E2E) speech recognition models that have surpassed traditional models (with separate acoustic, pronunciation, and language model components). To address the issue of inconsistency between the length of the speech sequence and the output sequence, end-to-end speech recognition techniques can be typically categorized as follows: connectionist temporal classification (CTC)~\cite{ctc_2006}, recurrent neural network transducer (RNN-T)~\cite{graves2012sequence}, and attention-based approaches~\cite{chorowski2015attention,bahdanau2016end,chan2016listen} are the most prominent approaches in this field. In recent years, models based on the Transformer~\cite{vaswani2017attention} structure have shown excellent performance across various tasks, including natural language understanding, machine translation, and speech recognition.
% streaming asr
RNN-T-based ASR systems have demonstrated advanced performance for streaming and online applications, with successful deployment in production systems~\cite{li2019improving}. Nevertheless, attention-based encoder-decoder architectures represent the most effective end-to-end ASR systems. However, due to the characteristics of the attention mechanism, it cannot be directly implemented in streaming scenarios, which makes its deployment in streaming systems still challenging and hinders its wide application in practice. To address this limitation, alternative streaming ASR approaches based on attention systems have been proposed, including Monotonic Attention~\cite{chiu2018monotonic,Merboldt2019AnAO,Inaguma2020EnhancingMM}, Chunk-wise~\cite{tsunoo2021streaming,tian2020synchronous}, Accumulation of Information~\cite{9383613} and Trigger Attention~\cite{moritz2019triggered,moritz2020streaming}, among others.

% multi-model
Multimodal data can aid audio information for ASR. In Project Aria~\cite{somasundaram2023project}, smart glasses equipped with five cameras (two Mono Scene, one RGB, and two Eye Tracking cameras) as well as non-visual sensors (two IMUs, magnetometer, barometer, GPS, Wi-Fi beacon, Bluetooth beacon, and microphones), can provide useful multi-modal information for speech recognition of the person wearing them.

In this paper, our streaming ASR system uses the Fast-Conformer~\cite{rekesh2023fast} architecture as an encoder of the RNN-T model. We leverage multi-overlap rates multi-channel simulated data and a 1:1 match training strategy between real and simulated data, to improve the generalizability of the model on real-world data. Besides, we incorporate the IMU unit data with audio data, to assist multi-modal ASR architecture for capturing the useful information embedded in IMU unit data for speech recognition. Experimental results show the efficiency of the proposed audio-only and multi-modal ASR model in streaming speech recognition in two-person conversation scenarios.

\section{System Description}
% beamformer+audio-cnn+fast-conformer+decoder rnnt-ctc decoder）
\subsection{Audio-only ASR Architecture}
\begin{figure}[t]
\centering
\includegraphics[width=0.27\textwidth]{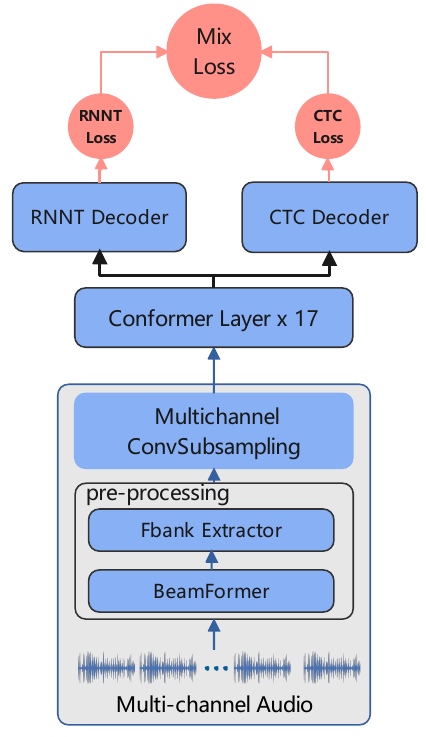}
\caption{The audio-only model Architecture for streaming ASR in smart glasses.}
\label{fig: ao-asr}
\end{figure}

We use the architecture shown in Fig. \ref{fig: ao-asr} as our audio-only ASR model, which consists of a front-end with multiple super-directive beamformers followed by an ASR module. Specifically, the N channels of raw audio data are fed into the beamformer front-end, called the Linearly Constrained Minimum Variance (NLCMV) beamforming technique~\cite{feng2023directional}. The multiple NLCMV beamformers pre-process the raw multi-channel audio into K horizontal steering directions around the smart-glasses devices plus one in the speaker's mouth direction. We use the predetermined beamformer weights with horizontal seeing directions K = 12, leading to 13-channel beamformed outputs. Then we extract the log-Mel features for each beamformer direction and concatenate them together. This concatenated vector constitutes the input of the ASR encoder. 

The ASR module is modified to receive multiple input streams from single-channel ASR systems. It is trained via serialized output training~\cite{kanda22b_interspeech} to detect speech from different directions, which allows it to classify and separate speech signals arriving from various directions, and leverage the differences in directional output from the beamformers. We use a Neural Transducer as our end-to-end ASR model, which consists of three components: an encoder, a prediction network, and a joint network. The audio encoder we adopt is a multi-channel convolutional downsampling module followed by a Fast-Conformer network ~\cite{rekesh2023fast}, which processes the input sequence and produces a sequence of acoustic representations. The Fast-Conformer can enhance the conformer architecture for efficient training and inference, leveraging a novel downsampling schema and replacing global attention with limited context attention. The prediction neural network acts as an internal language model or decoder to generate a representation. Lastly, the joiner network takes the output representations from the encoder and prediction network as input and creates the joint representation. The conventional hybrid architecture uses two decoders, one CTC decoder and one RNN-T decoder as shown in Fig. \ref{fig: ao-asr}. Both decoders share a single encoder and any of the two decoders can be used for inference after training is done. During training, we only use the RNN-T decoder to train our models and adopt the losses of the RNN-T decoder. Besides, we make all the convolution layers including those in the downsampling layers fully causal by padding of size $k - 1$ to the left of the input sequence where k is the convolution kernel size, and zero padding for the right side.

\subsection{Streaming Training Mechanism}
% 流式训练机制 (训练attention+推理chunk-cache）训练+推理机制
The model is non-autoregressive in the training stage, but the model of streaming ASR is not allowed to see the global information, so the system uses some methods to keep consistency in the training and inference process. Firstly, the system does not use normalization to process the Fbank features input to the Fast-Conformer model, while all the convolutional and downsampling layers of the model are modified to causal convolution. Moreover, the original BatchNorm \cite{ioffe2015batch} layers are replaced by LayerNorm~\cite{ba2016layer} layers to prevent seeing global information.

For the self-attention layers, the model uses the chunk-aware look-ahead approach~\cite{noroozi2024stateful} to limit the length of the context. This approach divides the audio input into chunks. The token of the same chunk can access all other tokens in the same chunk and a limited number of tokens from previous chunks. Unlike regular look-ahead, the effective look-ahead of the chunk-aware approach is independent of the depth of the model, which reduces unnecessary repetitive computation and accelerates the speed of inference. Meanwhile, the average look-ahead of each token in chunk-aware look-ahead is larger than that of regular look-ahead, which makes it possible to obtain better accuracy with the same delay.

During the inference stage, the model uses a caching mechanism to convert the non-autoregressive Fast-Conformer encoder into an autoregressive recurrent model by using a cache for activations computed from previous time steps. The cache size of the self-attention layers grows from zero to $L_c$, which is the past context size. For a self-attention layer with a left context length of $L_c$, the cache is empty in the first streaming step. For each subsequent streaming step, the number of activations from the input to self-attention layers is added to the cache, and old values are removed. The cache will finally grow to size $L_c$. For a model with batch size $B$, with L self-attention layers, and a hidden size D for each self-attention layer, a cache matrix of size $L \times B \times C_{\text{mha}} \times D$ is required, where $0 \leq C_{\text{mha}} \leq L_c$.

\subsection{Multi-channel Audio Simulation}
\label{ssec: mcas}
The multi-modal conversations in smart glasses (MMCSG) task~\cite{ZmolikovaEtAl2024} involves natural conversations between two people recorded with smart Aria glasses. The goal is to obtain speaker-attributed transcriptions in a streaming fashion, using audio, video, and inertial measurement unit (IMU) input modalities. We use Librispeech dataset~\cite{libri} as the raw conversation data and the Deep Noise Suppression (DNS) challenge noises dataset~\cite{dubey2023icassp} resampled to 16kHz as the noise data for simulation. In order to match the two-person dialog scenario of the MMCSG task, we randomly combine the single-channel raw audio into pairs, with one as SELF and the other as OTHER. Meanwhile, in order to satisfy the streaming needs, the original segment-level transcript will use the publicly available pre-computed alignment of Librispeech to accomplish the word-level alignment.
To simulate a real conversation, we first randomly select the overlap length of a pair of audios within a certain ratio. Then we copy the original single-channel audio and noise wave into seven channels and use real multi-channel SELF and OTHER speakers' room impulse response (RIR) information to convolve the expanded seven-channel audio to obtain the simulated seven-channel audio, as well as the noise data. Finally, we select the signal-noise ratio (SNR) of the noise in a weighted manner, adding to the simulated seven-channel audio and obtaining the simulated data.

Given the scarcity of real audio data, we generated nearly 2,000 hours of two-person dialogue data in simulated conversational scenarios, in contrast to the less than 10 hours of real data available. Training the model with a direct mix of simulated and real data poses a risk of overfitting the simulated data. This is because the real data can be overwhelmed by the vast amount of simulated data, and random selection during training might result in batches composed entirely of simulated data, which hinders the model's learning process. To mitigate this risk, we propose a balanced training strategy. In each training batch, for every real audio sample randomly selected, we randomly select a corresponding simulated audio sample from our simulated dataset. This approach ensures that each batch consists of 50$\%$ real audio data and 50$\%$ simulated audio data. Real data provides the model with authentic variations and noise characteristics present in natural environments, while simulated data offers a diverse and extensive set of training examples. By maintaining this balance, we aim to enhance the model’s ability to generalize from both real and simulated data, thereby reducing the risk of overfitting and improving overall performance. This balanced data strategy helps prevent the model from becoming biased towards the synthetic patterns present in the simulated data, thereby enhancing its ability to perform well on unseen real-world data.

\subsection{Multi-modal ASR Architecture}
\begin{figure}[t]
\centering
\includegraphics[width=0.35\textwidth]{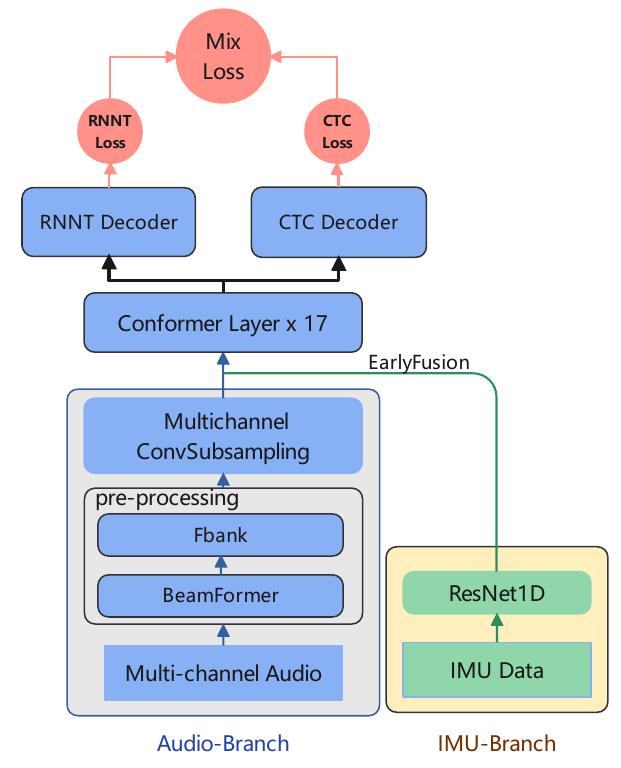}
\caption{The Multi-modal model Architecture for streaming ASR in smart glasses.}
\label{fig: mm-asr}
\end{figure}

Fig. \ref{fig: mm-asr} is our multi-modal architecture. To effectively utilize multi-modal data, we integrate the IMU data and audio data to bring improvements to the audio-only ASR model. The IMU unit, positioned on the glasses, can effectively capture the movements and speech-related vocalizations of the SELF speaker, thereby aiding the model's inference when combined with audio data. Given the IMU unit's sensitivity to different frequencies, we apply high-pass filtering to its data, removing components below 20 Hz, which typically contain high noise interference~\cite{gao2022inertiear}. This preprocessing step helps to eliminate irrelevant noise and retain useful motion information. We then use a 1D version of ResNet18 architecture to encode the IMU data~\cite{9196860}. 

The encoded IMU representations were then concatenated with downsampled audio features and fed into the Fast-Conformer encoder for learning. Meanwhile, the rest of the multi-modal ASR structure remains consistent with the single-audio model. This fusion allows the model to leverage both the motion data from the IMU and the audio data, potentially enhancing its overall performance. We conducted separate experiments with accelerometer and gyroscope data to determine the auxiliary performance of each modality in conjunction with audio. These experiments aimed to identify the effective way to integrate multi-modal data to support audio-based tasks.

\subsection{Optimization and Loss}
All models are optimized using Adam with $\beta_1 = 0.9$, $\beta_2=0.98$, and an initial learning rate of $0.5$. The learning rate is warmed up linearly in the first 5000 steps and decreases proportionally to the inverse square root of the step number. And we only use RNN-T loss during training steps.

\section{Experiments}
\subsection{Experimental Setups}
The MMCSG dataset is a multi-modal dataset of in-room dialog scenes recorded using Project Aria glasses. The dataset consists of 172 recordings for training, 169 recordings for development, and 189 recordings for evaluation. Each recording contains a conversation between two participants with optional background noise. Both conversation participants wore Aria glasses to record the conversation. The wearer of the glasses is referred to as SELF and the conversation partner is referred to as OTHER. the room was also interrupted by the moderator, who appeared in some of the video recordings but did not speak.
Each recording in the dataset includes the following modalities: seven channels of audio recorded at 48 kHz, video with facial blurring recorded at 15 fps, 720x720 resolution, and 8 bpp, and accelerometer and gyroscope signals recorded at a sampling rate of 1 kHz using a BMI085 IMU device.
In addition to using data from the MMCSG dataset, we simulated 2000 hours of data for model training using all 960 hours of data from the Librispeech dataset as the audio of the conversation between SELF and OTHER and selecting the DNS noise dataset as the noise data. All the simulated data are as allowed by the rules and in both the audio-only and multi-modal systems, we utilize a Fast-Conformer pre-trained model \cite{nvidia2023fastconformerASR} to initialize the network.

We train the streaming models at delay thresholds of 1000 ms, 350 ms, and 150 ms by calculating the average delay time for each system to correctly recognize words according to the task evaluation rules, and use the average multi-speaker WER of SELF and OTHER speakers to select the models on the dev dataset, where the multi-speaker WER includes errors in recognizing speakers in addition to the standard insertion, deletion, and substitution errors.

\subsection{Audio-only Experiments and Analysis}
\begin{table}[t]
	\renewcommand\arraystretch{1.25}
	\newcolumntype{L}[1]{>{\raggedright\arraybackslash}p{#1}}
	\newcolumntype{C}[1]{>{\centering\arraybackslash}p{#1}}
	\newcolumntype{R}[1]{>{\raggedleft\arraybackslash}p{#1}}
	\centering
	\caption{Directly mixing simulated data with real data for training. `+' denotes that the current dataset is added to the previous line.}
        \vspace{-0.5cm}
	\label{tab: Simu}\medskip
	\resizebox{8 cm}{!}{\begin{tabular}{c|c|c|c|c}
		\toprule[1 pt]
		Dataset & Latency & SELF & OTHER & OVERALL \\
            Duration & Mean [s] & WER [$\%$] & WER [$\%$] & WER [$\%$] \\
		\midrule
            \multirow{3}{*}{\shortstack{Real Data\\ (7h)}}           & 0.150 & 17.9 & 24.4 & 21.15 \\
                                                                    & 0.340 & 15.0 & 21.4 & 18.20 \\
                                                                    & 0.620 & 14.3 & 20.3 & 17.30 \\
		\midrule
            \multirow{3}{*}{\shortstack{+ Speed Perturb Data\\ (14h)}} & 0.124 & 18.3 & 23.3 & 20.80 \\
                                                                    & 0.329 & 15.1 & 20.4 & 17.75 \\
                                                                    & 0.619 & 14.1 & 19.6 & 16.85 \\
		\midrule
            \multirow{3}{*}{\shortstack{+ Simulated Data\\ (1200h)}} & 0.136 & 14.6 & 22.0 & 18.30 \\
                                                                    & 0.332 & 12.4 & 19.7 & 16.05 \\
                                                                    & 0.619 & 11.7 & 18.9 & 15.30 \\
		\midrule
            \multirow{3}{*}{\shortstack{+ Multi-overlap Simulated Data\\ (1911h)}} & 0.144 & 13.7 & 21.5 & 17.60 \\
                                                                    & 0.332 & 11.4 & 19.2 & 15.30 \\
                                                                    & 0.619 & 10.8 & 18.4 & 14.60 \\
            \bottomrule[1 pt]
	\end{tabular}}
\end{table}

\begin{table}[t]
	\renewcommand\arraystretch{1.25}
	\newcolumntype{L}[1]{>{\raggedright\arraybackslash}p{#1}}
	\newcolumntype{C}[1]{>{\centering\arraybackslash}p{#1}}
	\newcolumntype{R}[1]{>{\raggedleft\arraybackslash}p{#1}}
	\centering
	\caption{Train with a 1:1 match between real and simulated data.}
        \vspace{-0.5cm}
	\label{tab: Simu-strategy}\medskip
	\resizebox{8 cm}{!}{\begin{tabular}{c|c|c|c|c}
		\toprule[1 pt]
		Dataset & Latency & SELF & OTHER & OVERALL \\
            Duration & Mean [s] & WER [$\%$] & WER [$\%$] & WER [$\%$] \\
		\midrule
            \multirow{3}{*}{\shortstack{1 fold Real + simulated data\\ (14h)}} & 0.160 & 16.1 & 23.7 & 19.90 \\
                                                                    & 0.351 & 13.7 & 20.9 & 17.30 \\
                                                                    & 0.635 & 12.9 & 20.0 & 16.45 \\
		\midrule
            \multirow{3}{*}{\shortstack{80 folds Real + simulated data\\ (1120h)}} & 0.181 & 12.5 & 19.6 & 16.05 \\
                                                                    & 0.362 & 11.3 & 18.2 & 14.75 \\
                                                                    & 0.645 & 10.9 & 17.7 & 14.30 \\
            \bottomrule[1 pt]
	\end{tabular}}
\end{table}

\begin{table}[t]
	\renewcommand\arraystretch{1.25}
	\newcolumntype{L}[1]{>{\raggedright\arraybackslash}p{#1}}
	\newcolumntype{C}[1]{>{\centering\arraybackslash}p{#1}}
	\newcolumntype{R}[1]{>{\raggedleft\arraybackslash}p{#1}}
	\centering
	\caption{Experiments on the simulation dataset with the addition of visible context information for Fast-Conformer.}
        \vspace{-0.5cm}
	\label{tab: attn}\medskip
	\resizebox{8 cm}{!}{\begin{tabular}{c|c|c|c|c}
		\toprule[1 pt]
		\multirow{2}{*}{\shortstack{Attention\\Context Size}} & Latency & SELF & OTHER & OVERALL \\
                                   & Mean [s] & WER [$\%$] & WER [$\%$] & WER [$\%$] \\
		\midrule
            $[$84, 20$]$ & 0.871 & 10.3 & 17.7 & 14.00 \\
            \bottomrule[1 pt]
	\end{tabular}}
\end{table}

We conducted ablation experiments with augmented multi-channel data. We first applied data augmentation (e.g., speed perturb) to increase the diversity of real data, depicted as `+ Speed Perturb Data' in Table \ref{tab: Simu}. Using the RTTM files in the training set, we computed and found that the overlap rate is as high as 11$\%$ or more in the two-person conversation scenario of the MMCSG dataset. Therefore, we adjusted the overlap rate of the simulation data to be higher and generated simulation data with multiple overlap rates with pluralistic overlap rates from 5$\%$ to 30$\%$ to cope with the realistic situation. Mixing real audio data, real audio data after perturbation enhancement, and these simulation data with multiple overlap rates, we conducted experiments as shown in the table, and the increase in the amount of simulation data brought optimization to the WER metrics. However, when the simulation data reaches a certain amount of data, the real audio will be submerged in the simulation data, and the performance of the system is no longer improved. Therefore, we adopt the simulation training strategy as described in Section \ref{ssec: mcas} for further experiments in Table \ref{tab: Simu-strategy}. It can be seen that the latter approach leads to a greater performance improvement for a close amount of training data. However, it exceeds the specified low-latency threshold under the same inference conditions for the same visible future context.

As shown in Table~\ref{tab: attn}, we adjust the mask for the attention module in the Fast-Conformer model so that the attention module of the model has a longer visible context to get better recognition performance at higher latency.

\subsection{Multi-modal Experiments and Analysis}
\vspace{-0.2cm}
\begin{table}[h]
	\renewcommand\arraystretch{1}
	\newcolumntype{L}[1]{>{\raggedright\arraybackslash}p{#1}}
	\newcolumntype{C}[1]{>{\centering\arraybackslash}p{#1}}
	\newcolumntype{R}[1]{>{\raggedleft\arraybackslash}p{#1}}
	\centering
	\caption{Trained with real audio, accelerometer, and gyroscope data, using ResNet18-1D as the feature extraction module.}
        \vspace{-0.5cm}
	\label{tab: multi_imu}\medskip
	\resizebox{8 cm}{!}{\begin{tabular}{c|c|c|c|c}
		\toprule[1 pt]
		Dataset & Latency & SELF & OTHER & OVERALL \\
            Modality & Mean [s] & WER [$\%$] & WER [$\%$] & WER [$\%$] \\

            \midrule
            \multirow{3}{*}{\shortstack{Audio-only \\ Baseline}}           & 0.150 & 17.9 & 24.4 & 21.15 \\
                                                                    & 0.340 & 15.0 & 21.4 & 18.20 \\
                                                                    & 0.620 & 14.3 & 20.3 & 17.30 \\
		\midrule
            \multirow{3}{*}{\shortstack{Audio \\ + Accelerometer}}           & 0.125 & 18.8 & 22.9 & 20.85 \\
                                                                    & 0.331 & 15.5 & 19.9 & 17.50 \\
                                                                    & 0.617 & 14.0 & 19.1 & 16.55 \\
		\midrule
            \multirow{3}{*}{\shortstack{Audio \\ + Gyroscope}} & 0.136 & 18.1 & 22.8 & 20.45 \\
                                                                    & 0.342 & 15.1 & 20.1 & 17.60 \\
                                                                    & 0.620 & 14.1 & 19.2 & 16.65 \\
		\midrule
            \multirow{3}{*}{\shortstack{Audio \\ + 6-axis IMU }} & 0.139 & 17.6 & 23.2 & 20.40 \\
                                                                    & 0.347 & 14.6 & 19.9 & 17.25 \\
                                                                    & 0.658 & 13.8 & 19.1 & 16.45 \\
		\midrule
            \multirow{3}{*}{\shortstack{Audio \\ + 6-axis IMU \\ + w/o Filter }} & 0.144 & 17.9 & 24.3 & 21.10 \\
                                                                    & 0.350 & 15.1 & 21.1 & 18.10 \\
                                                                    & 0.628 & 14.1 & 20.1 & 17.10 \\
            \bottomrule[1 pt]
	\end{tabular}}
\end{table}
The multi-modal system was not submitted since the experimental results were not excellent but we still show some exploratory experiments we did in the multi-modal system. 
According to~\cite{zhang2015accelword,michalevsky2014gyrophone}, sound vibrations can affect the accelerometer and gyroscope, allowing the IMU device to capture the acoustic fundamental frequency information of the wearer. This can provide more information to help the model in the MMCSG task to supplement the information of the SELF speaker during the overlap segment, distinguish between the SELF and OTHER speakers, and improve the performance of the model.
We applied high-pass filtering to the IMU data to remove components below 20 Hz. This is because the wearer's head movements and body posture introduce interference to the IMU recordings, with most of this interference occurring below 20 Hz. The filtered data was then fed into a ResNet18-1D and performed early fusion with audio features extracted through a downsampling network.
As shown in Tabel~\ref{tab: multi_imu}, the results of our system using all six-axis data from the IMU are slightly better than the baseline, while the network becomes less effective if only accelerometer or gyroscope data are used. This indicates that valuable information is contained in both the accelerometer and gyroscope. For unfiltered IMU data, the result is almost the same as the baseline, indicating that the original IMU data is noisy and difficult for the network to learn.

\subsection{The Latency and Wer in Final Submission Systems}

\begin{table}[t]
	\renewcommand\arraystretch{1.25}
	\newcolumntype{L}[1]{>{\raggedright\arraybackslash}p{#1}}
	\newcolumntype{C}[1]{>{\centering\arraybackslash}p{#1}}
	\newcolumntype{R}[1]{>{\raggedleft\arraybackslash}p{#1}}
	\centering
	\caption{The latency and wer on the dev dataset of the final submission systems.}
        \vspace{-0.5cm}
	\label{tab: submission}\medskip
	\resizebox{8 cm}{!}{\begin{tabular}{c|c|c|c|c|c}
		\toprule[1 pt]
		System & Latency & Attention & SELF & OTHER & OVERALL \\
            Number & Mean [s] & Context Size & WER [$\%$] & WER [$\%$] & WER [$\%$] \\
		\midrule
            1 & 0.130 & $[$70, 1$]$ & 14.0 & 21.3 & 17.65 \\
            2 & 0.126 & $[$70, 1$]$ & 13.6 & 21.7 & 17.65 \\    
            3 & 0.144 & $[$70, 1$]$ & 13.7 & 21.5 & 17.60 \\
		\midrule
            4 & 0.254 & $[$70, 4$]$ & 11.8 & 19.9 & 15.85 \\
            5 & 0.323 & $[$70, 6$]$ & 11.4 & 19.3 & 15.35 \\
            6 & 0.332 & $[$70, 6$]$ & 11.4 & 19.2 & 15.30 \\
		\midrule
            7 & 0.645 & $[$70, 13$]$ & 10.9 & 17.7 & 14.30 \\
            8 & 0.964 & $[$88, 21$]$ & 10.4 & 18.1 & 14.25 \\
            9 & 0.871 & $[$84, 20$]$ & 10.3 & 17.7 & 14.00 \\
		\midrule
            10 & - & - & 9.9 & 15.4 & 12.65 \\
            11 & - & - & 8.6 & 15.7 & 12.15 \\
            \bottomrule[1 pt]
	\end{tabular}}
\end{table}

\begin{table}[t]
	\renewcommand\arraystretch{1.25}
	\newcolumntype{L}[1]{>{\raggedright\arraybackslash}p{#1}}
	\newcolumntype{C}[1]{>{\centering\arraybackslash}p{#1}}
	\newcolumntype{R}[1]{>{\raggedleft\arraybackslash}p{#1}}
	\centering
	\caption{The latency and wer on the eval dataset of the final submission systems.}
        \vspace{-0.5cm}
	\label{tab: submission_eval}\medskip
	\resizebox{8 cm}{!}{\begin{tabular}{c|c|c|c|c|c}
		\toprule[1 pt]
		System & Latency & Attention & SELF & OTHER & OVERALL \\
            Number & Mean [s] & Context Size & WER [$\%$] & WER [$\%$] & WER [$\%$] \\
		\midrule
            1 & 0.137 & $[$70, 1$]$ & 13.6 & 22.6 & 18.10 \\
            2 & 0.126 & $[$70, 1$]$ & 13.0 & 22.9 & 17.95 \\    
            3 & 0.142 & $[$70, 1$]$ & 13.4 & 23.0 & 18.20 \\
		\midrule
            4 & 0.257 & $[$70, 4$]$ & 11.5 & 21.1 & 16.30 \\
            5 & 0.317 & $[$70, 6$]$ & 11.1 & 20.5 & 15.80 \\
            6 & 0.332 & $[$70, 6$]$ & 11.2 & 20.3 & 15.75 \\
		\midrule
            7 & 0.637 & $[$70, 13$]$ & 10.7 & 19.0 & 14.85 \\
            8 & 0.977 & $[$88, 21$]$ & 10.2 & 19.1 & 14.65 \\
            9 & 0.879 & $[$84, 20$]$ & 9.9 & 18.9 & 14.40 \\
		\midrule
            10 & - & - & 9.7 & 16.1 & 12.90 \\
            11 & - & - & 8.4 & 16.7 & 12.55 \\
            \bottomrule[1 pt]
	\end{tabular}}
\end{table}

We submitted 11 systems spread over a range of latency thresholds of $<$150ms, 150$\sim$350ms, 350ms$\sim$1000ms, and $>$1000ms. We tested their performance of average system latency and wer on the dev dataset as listed in Table \ref{tab: submission}. In terms of latency, the system only changes the att\_context\_size settings, all the rest of the configuration is the same as the baseline, and the specific changes are shown in Table~\ref{tab: submission}. Systems 1-3 are in the below 150ms latency category, systems 4-6 are in the 150$\sim$350ms latency category, and systems 7-9 are in the 350$\sim$1000ms latency category. The systems 10 and 11 are the non-streaming systems by replacing the context-aware attention mask in the ASR encoder module with the global-aware attention mask, thus we did not report the latency on the non-streaming systems.

In Table~\ref{tab: submission_eval}, we show the results of our submission system on the final eval test set, where system 10 came in first place in the track with a latency of $>$1000ms.

\bibliography{mybib}

\begin{thebibliography}{10}
\providecommand{\url}[1]{#1}
\csname url@samestyle\endcsname
\providecommand{\newblock}{\relax}
\providecommand{\bibinfo}[2]{#2}
\providecommand{\BIBentrySTDinterwordspacing}{\spaceskip=0pt\relax}
\providecommand{\BIBentryALTinterwordstretchfactor}{4}
\providecommand{\BIBentryALTinterwordspacing}{\spaceskip=\fontdimen2\font plus
\BIBentryALTinterwordstretchfactor\fontdimen3\font minus
  \fontdimen4\font\relax}
\providecommand{\BIBforeignlanguage}[2]{{%
\expandafter\ifx\csname l@#1\endcsname\relax
\typeout{** WARNING: IEEEtran.bst: No hyphenation pattern has been}%
\typeout{** loaded for the language `#1'. Using the pattern for}%
\typeout{** the default language instead.}%
\else
\language=\csname l@#1\endcsname
\fi
#2}}
\providecommand{\BIBdecl}{\relax}
\BIBdecl

\bibitem{li2022recent}
J.~Li \emph{et~al.}, ``Recent advances in end-to-end automatic speech
  recognition,'' \emph{APSIPA Transactions on Signal and Information
  Processing}, vol.~11, no.~1, 2022.

\bibitem{ctc_2006}
A.~Graves, S.~Fern{\'a}ndez, F.~Gomez, and J.~Schmidhuber, ``Connectionist
  temporal classification: labelling unsegmented sequence data with recurrent
  neural networks,'' in \emph{Proceedings of the 23rd international conference
  on Machine learning}, 2006, pp. 369--376.

\bibitem{graves2012sequence}
A.~Graves, ``Sequence transduction with recurrent neural networks,''
  \emph{arXiv preprint arXiv:1211.3711}, 2012.

\bibitem{chorowski2015attention}
J.~K. Chorowski, D.~Bahdanau, D.~Serdyuk, K.~Cho, and Y.~Bengio,
  ``Attention-based models for speech recognition,'' \emph{Advances in neural
  information processing systems}, vol.~28, 2015.

\bibitem{bahdanau2016end}
D.~Bahdanau, J.~Chorowski, D.~Serdyuk, P.~Brakel, and Y.~Bengio, ``End-to-end
  attention-based large vocabulary speech recognition,'' in \emph{2016 IEEE
  international conference on acoustics, speech and signal processing
  (ICASSP)}.\hskip 1em plus 0.5em minus 0.4em\relax IEEE, 2016, pp. 4945--4949.

\bibitem{chan2016listen}
W.~Chan, N.~Jaitly, Q.~Le, and O.~Vinyals, ``Listen, attend and spell: A neural
  network for large vocabulary conversational speech recognition,'' in
  \emph{2016 IEEE international conference on acoustics, speech and signal
  processing (ICASSP)}.\hskip 1em plus 0.5em minus 0.4em\relax IEEE, 2016, pp.
  4960--4964.

\bibitem{vaswani2017attention}
A.~Vaswani, N.~Shazeer, N.~Parmar, J.~Uszkoreit, L.~Jones, A.~N. Gomez,
  {\L}.~Kaiser, and I.~Polosukhin, ``Attention is all you need,''
  \emph{Advances in neural information processing systems}, vol.~30, 2017.

\bibitem{li2019improving}
J.~Li, R.~Zhao, H.~Hu, and Y.~Gong, ``Improving rnn transducer modeling for
  end-to-end speech recognition,'' in \emph{2019 IEEE Automatic Speech
  Recognition and Understanding Workshop (ASRU)}.\hskip 1em plus 0.5em minus
  0.4em\relax IEEE, 2019, pp. 114--121.

\bibitem{chiu2018monotonic}
\BIBentryALTinterwordspacing
C.-C. Chiu* and C.~Raffel*, ``Monotonic chunkwise attention,'' in
  \emph{International Conference on Learning Representations}, 2018. [Online].
  Available: \url{https://openreview.net/forum?id=Hko85plCW}
\BIBentrySTDinterwordspacing

\bibitem{Merboldt2019AnAO}
\BIBentryALTinterwordspacing
A.~Merboldt, A.~Zeyer, R.~Schl{\"u}ter, and H.~Ney, ``An analysis of local
  monotonic attention variants,'' in \emph{Interspeech}, 2019. [Online].
  Available: \url{https://api.semanticscholar.org/CorpusID:203144513}
\BIBentrySTDinterwordspacing

\bibitem{Inaguma2020EnhancingMM}
\BIBentryALTinterwordspacing
H.~Inaguma, M.~Mimura, and T.~Kawahara, ``Enhancing monotonic multihead
  attention for streaming asr,'' in \emph{Interspeech}, 2020. [Online].
  Available: \url{https://api.semanticscholar.org/CorpusID:218684429}
\BIBentrySTDinterwordspacing

\bibitem{tsunoo2021streaming}
E.~Tsunoo, Y.~Kashiwagi, and S.~Watanabe, ``Streaming transformer asr with
  blockwise synchronous beam search,'' in \emph{2021 IEEE Spoken Language
  Technology Workshop (SLT)}.\hskip 1em plus 0.5em minus 0.4em\relax IEEE,
  2021, pp. 22--29.

\bibitem{tian2020synchronous}
Z.~Tian, J.~Yi, Y.~Bai, J.~Tao, S.~Zhang, and Z.~Wen, ``Synchronous
  transformers for end-to-end speech recognition,'' in \emph{ICASSP 2020-2020
  IEEE International Conference on Acoustics, Speech and Signal Processing
  (ICASSP)}.\hskip 1em plus 0.5em minus 0.4em\relax IEEE, 2020, pp. 7884--7888.

\bibitem{9383613}
M.~Li, C.~Zorilă, and R.~Doddipatla, ``Transformer-based online speech
  recognition with decoder-end adaptive computation steps,'' in \emph{2021 IEEE
  Spoken Language Technology Workshop (SLT)}, 2021, pp. 1--7.

\bibitem{moritz2019triggered}
N.~Moritz, T.~Hori, and J.~Le~Roux, ``Triggered attention for end-to-end speech
  recognition,'' in \emph{ICASSP 2019-2019 IEEE International Conference on
  Acoustics, Speech and Signal Processing (ICASSP)}.\hskip 1em plus 0.5em minus
  0.4em\relax IEEE, 2019, pp. 5666--5670.

\bibitem{moritz2020streaming}
N.~Moritz, T.~Hori, and J.~Le, ``Streaming automatic speech recognition with
  the transformer model,'' in \emph{ICASSP 2020-2020 IEEE International
  Conference on Acoustics, Speech and Signal Processing (ICASSP)}.\hskip 1em
  plus 0.5em minus 0.4em\relax IEEE, 2020, pp. 6074--6078.

\bibitem{somasundaram2023project}
K.~Somasundaram, J.~Dong, H.~Tang, J.~Straub, M.~Yan, M.~Goesele, J.~J. Engel,
  R.~De~Nardi, and R.~Newcombe, ``Project aria: A new tool for egocentric
  multi-modal ai research,'' \emph{arXiv preprint arXiv:2308.13561}, 2023.

\bibitem{rekesh2023fast}
D.~Rekesh, N.~R. Koluguri, S.~Kriman, S.~Majumdar, V.~Noroozi, H.~Huang,
  O.~Hrinchuk, K.~Puvvada, A.~Kumar, J.~Balam \emph{et~al.}, ``Fast conformer
  with linearly scalable attention for efficient speech recognition,'' in
  \emph{2023 IEEE Automatic Speech Recognition and Understanding Workshop
  (ASRU)}.\hskip 1em plus 0.5em minus 0.4em\relax IEEE, 2023, pp. 1--8.

\bibitem{feng2023directional}
T.~Feng, J.~Lin, Y.~Huang, W.~He, K.~Kalgaonkar, N.~Moritz, L.~Wan, X.~Lei,
  M.~Sun, and F.~Seide, ``Directional source separation for robust speech
  recognition on smart glasses,'' \emph{arXiv preprint arXiv:2309.10993}, 2023.

\bibitem{kanda22b_interspeech}
N.~Kanda, J.~Wu, Y.~Wu, X.~Xiao, Z.~Meng, X.~Wang, Y.~Gaur, Z.~Chen, J.~Li, and
  T.~Yoshioka, ``{Streaming Speaker-Attributed ASR with Token-Level Speaker
  Embeddings},'' in \emph{Proc. Interspeech 2022}, 2022, pp. 521--525.

\bibitem{ioffe2015batch}
S.~Ioffe and C.~Szegedy, ``Batch normalization: Accelerating deep network
  training by reducing internal covariate shift,'' in \emph{International
  conference on machine learning}.\hskip 1em plus 0.5em minus 0.4em\relax pmlr,
  2015, pp. 448--456.

\bibitem{ba2016layer}
J.~L. Ba, J.~R. Kiros, and G.~E. Hinton, ``Layer normalization,'' \emph{arXiv
  preprint arXiv:1607.06450}, 2016.

\bibitem{noroozi2024stateful}
V.~Noroozi, S.~Majumdar, A.~Kumar, J.~Balam, and B.~Ginsburg, ``Stateful
  conformer with cache-based inference for streaming automatic speech
  recognition,'' in \emph{ICASSP 2024-2024 IEEE International Conference on
  Acoustics, Speech and Signal Processing (ICASSP)}.\hskip 1em plus 0.5em minus
  0.4em\relax IEEE, 2024, pp. 12\,041--12\,045.

\bibitem{ZmolikovaEtAl2024}
K.~Žmolíková, S.~Merello, K.~Kalgaonkar, J.~Lin, N.~Moritz, P.~Ma, M.~Sun,
  H.~Chen, A.~Saliou, S.~Petridis, C.~Fuegen, and M.~Mandel, ``The chime-8
  mmcsg challenge: Multi-modal conversations in smart glasses,'' in \emph{CHiME
  Workshop on Speech Processing in Everyday Environments}, 2024.

\bibitem{libri}
V.~Panayotov, G.~Chen, D.~Povey, and S.~Khudanpur, ``Librispeech: an asr corpus
  based on public domain audio books,'' in \emph{2015 IEEE international
  conference on acoustics, speech and signal processing (ICASSP)}.\hskip 1em
  plus 0.5em minus 0.4em\relax IEEE, 2015, pp. 5206--5210.

\bibitem{dubey2023icassp}
H.~Dubey, A.~Aazami, V.~Gopal, B.~Naderi, S.~Braun, R.~Cutler, H.~Gamper,
  M.~Golestaneh, and R.~Aichner, ``Icassp 2023 deep noise suppression
  challenge,'' in \emph{ICASSP}, 2023.

\bibitem{gao2022inertiear}
M.~Gao, Y.~Liu, Y.~Chen, Y.~Li, Z.~Ba, X.~Xu, and J.~Han, ``Inertiear:
  Automatic and device-independent imu-based eavesdropping on smartphones,'' in
  \emph{IEEE INFOCOM 2022-IEEE Conference on Computer Communications}.\hskip
  1em plus 0.5em minus 0.4em\relax IEEE, 2022, pp. 1129--1138.

\bibitem{9196860}
S.~Herath, H.~Yan, and Y.~Furukawa, ``Ronin: Robust neural inertial navigation
  in the wild: Benchmark, evaluations, \& new methods,'' in \emph{2020 IEEE
  International Conference on Robotics and Automation (ICRA)}, 2020, pp.
  3146--3152.

\bibitem{nvidia2023fastconformerASR}
NVIDIA-NeMo, ``\textit{FastConformer Hybrid Large Streaming Multi (en-US)},''
  2023,
  \url{https://huggingface.co/nvidia/stt_en_fastconformer_hybrid_large_streaming_multi}.

\bibitem{zhang2015accelword}
L.~Zhang, P.~H. Pathak, M.~Wu, Y.~Zhao, and P.~Mohapatra, ``Accelword: Energy
  efficient hotword detection through accelerometer,'' in \emph{Proceedings of
  the 13th Annual International Conference on Mobile Systems, Applications, and
  Services}, 2015, pp. 301--315.

\bibitem{michalevsky2014gyrophone}
Y.~Michalevsky, D.~Boneh, and G.~Nakibly, ``Gyrophone: Recognizing speech from
  gyroscope signals,'' in \emph{23rd USENIX Security Symposium (USENIX Security
  14)}, 2014, pp. 1053--1067.

\end{thebibliography}
\bibliographystyle{IEEEtran}

\end{document}